\documentclass[12pt]{iopart}
\usepackage{iopams}  
\usepackage{graphicx}
\begin{document}

\title[A strontium lattice clock with $3\times 10^{-17}$ inaccuracy and its frequency]{A strontium lattice clock with $3\times 10^{-17}$ inaccuracy\\ and its frequency}

\author{Stephan~Falke, Nathan~Lemke, Christian~Grebing, Burghard~Lipphardt, Stefan~Weyers, Vladislav~Gerginov, Nils~Huntemann, Christian~Hagemann, Ali~Al-Masoudi, Sebastian~H{\"a}fner, Stefan~Vogt, Uwe~Sterr and Christian~Lisdat}

\address{Physikalisch-Technische Bundesanstalt (PTB);
Bundesallee 100;
38116 Braunschweig;
Germany}
\ead{christian.lisdat@ptb.de}
\begin{abstract}
We have measured the absolute frequency of the optical lattice clock based on $^{87}$Sr at PTB with an uncertainty of $3.9\times 10^{-16}$ using two caesium fountain clocks. This is close to the accuracy of today's best realizations of the SI second. The absolute frequency of the 5s$^2$~$^1$S$_0$~--~5s5p~$^3$P$_0$ transition in $^{87}$Sr is $429\,228\,004\,229\,873$.13(17)~Hz. Our result is in excellent agreement with recent measurements performed in different laboratories worldwide. We improved the total systematic uncertainty of our Sr frequency standard by a factor of five and reach $3\times 10^{-17}$, opening new prospects for frequency ratio measurements between optical clocks for fundamental research, geodesy, or optical clock evaluation.
\end{abstract}

\maketitle

\section{Introduction}
Motivated by the prospect for a redefinition of the SI second based on an optical transition, optical clocks employing various atomic species and reference transitions are currently being developed and refined worldwide~\cite{ros08,aka08,lem09,hun12,mad12,mcf12}. Although optical clocks achieve accuracies and stabilities superior to primary standards, only two frequency measurements limited by today's most accurate realization of the SI second have been reported so far~\cite{let13,tam14}. Such measurements are essential prerequisites for a possible redefinition of the second. Here, we present a third one that is in very good agreement with the measurement in Paris \cite{let13}. In addition to their role as future primary standards, optical clocks are of interest as sensitive probes in areas such as relativistic geodesy~\cite{bje85}, tests of relativity~\cite{sch09,cho10a}, and searches for variations of fundamental constants~\cite{ros08,pei04,for07,bla08,pei10}.

Two types of optical clocks have emerged: single ion clocks and optical lattice clocks. While in ion clocks a single reference atom is trapped utilizing its electric charge, in optical lattice clocks thousands of neutral atoms are trapped in an optical potential that is carefully tuned to the AC-Stark-shift-cancelling (magic) wavelength~\cite{kat03}. Several ion clock systems have demonstrated very low systematic uncertainties on the order of $10^{-17}$~\cite{ros08,hun12,mad12,cho10}. While the frequency stability of optical lattice clocks is superior~\cite{nic12,hin13}, their systematic uncertainty had been limited to approximately $10^{-16}$~\cite{lem09,let13,lud08,fal11}. However, a recent study placed strontium lattice clocks into the accuracy regime of the best ion clocks~\cite{blo13}.

The leading uncertainty contribution in lattice clocks is typically from the blackbody radiation (BBR) Stark shift~\cite{por06,mid11}. For the cases of Sr and Yb, recent measurements and calculations of the relevant atomic polarisabilities~\cite{she12,mid12a,saf12,bel12,saf13} have mitigated this key source of error, enabling lattice clocks to push the total systematic uncertainty below $10^{-16}$ for the first time. Simultaneously, several groups have reduced the noise of the interrogation oscillators~\cite{jia11,kes12a}, which has in turn led to significantly reduced clock instabilities~\cite{nic12,hin13,hag13}; this improved instability then allows for a very rapid and thorough investigation of systematic effects.

The accuracy of today's best Cs fountain clocks of $2-4\times 10^{-16}$~\cite{hea05,li11a,wey11,gue12a,Circular_T_Cs} sets a limit for the achievable accuracy of frequency measurements. For a measurement at this level the optical standard needs to be sufficiently accurate as well as operationally robust in order to accommodate the long averaging times required by the microwave standard. Moreover, the link between the optical and microwave standards must be carefully controlled to avoid possible errors. In this paper we report a frequency measurement of the clock transition 5s$^2$~$^1$S$_0$~--~5s5p~$^3$P$_0$ in $^{87}$Sr against two caesium fountain clocks at PTB with a total uncertainty of $3.9 \times 10^{-16}$. The systematic uncertainty of the strontium clock is below that of the caesium fountains by nearly one order of magnitude, benefiting from the much higher operating frequency of 429~THz compared to 9.2~GHz. This evaluation of a lattice clock accuracy is among the best~\cite{blo13} and indicates that the quest for the most accurate clock is open to atoms and ions.

The paper is organized as follows: First, we describe our strontium lattice clock setup. In Sec.~\ref{sec:srEffects} we report on the study of systematic frequency shifts and obtain the uncertainty budget. This is followed in Sec.~\ref{sec:absfreq} by a description of the absolute frequency measurement. Finally, we report and discuss our results in Sec.~\ref{sec:results}.

\begin{figure}
\centering
\includegraphics[width=1.0\columnwidth]{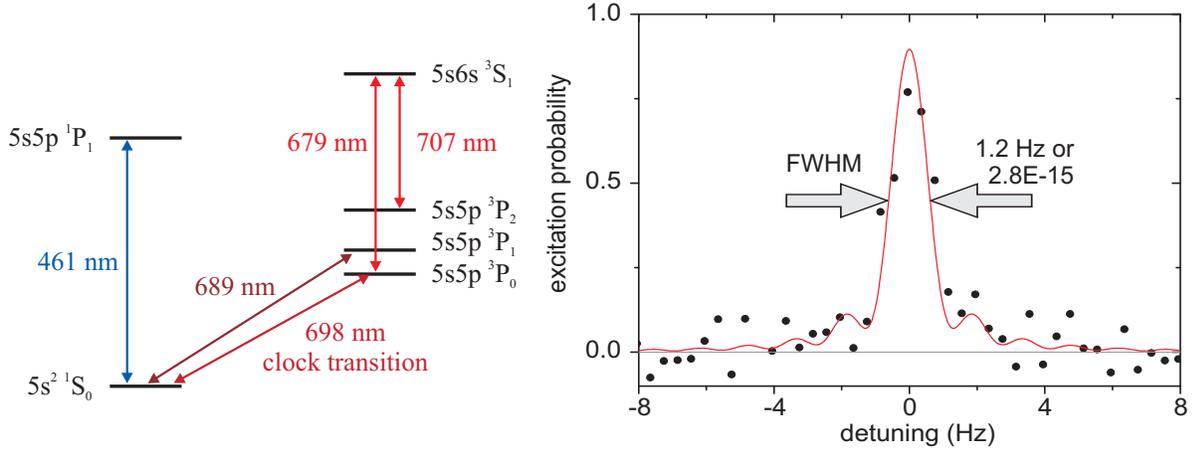}
\caption{Simplified level scheme of strontium with arrows indicating the required cooling and spectroscopy lasers and their wavelengths (left) and scan across the clock transition (right). The observed data (dots, single recording per point) are accompanied by a simulated spectrum (line) obtained by a numerical evaluation of the density matrix with a clock laser of linewidth of 0.08~Hz. The clock pulse length is 720~ms in both experiment and simulation. Due to noise and background subtraction, unphysical negative probabilities appear.}
\label{fig:scheme}
\end{figure}

\section{Strontium lattice clock setup}
\label{sec:prep}
The preparation of ultracold $^{87}$Sr atoms (nuclear spin $I = 9/2$) in an optical lattice follows commonly applied procedures. Here we give a general overview of the setup and refer to previous publications~\cite{fal11,lis09,mid12a} for details. A level scheme reduced to the relevant levels and transitions is shown in Fig.~\ref{fig:scheme}. The atom source is an oven operated at 770~K with no direct line of sight between the atom source and the magneto-optical trap (MOT). Atoms are Zeeman-slowed and captured in a MOT operated near the strong singlet transition $^1$S$_0$~--~$^1$P$_1$ at 461~nm to obtain samples at a few millikelvin. Originally an optical molasses had been applied to guide the slowed atoms into the MOT region~\cite{lis09} but it was deemed not necessary for efficient loading and is not present in this work. A second stage MOT operated on the $^1$S$_0$~--~$^3$P$_1$ transition at 689~nm cools the sample to about 2~$\mu$K. Here, two frequencies are applied to avoid population trapping by connecting the $^1$S$_0$ state to the $F=9/2$ and $F=11/2$ states of $^3$P$_1$~\cite{muk03}. Initially, this MOT uses high laser intensity and a laser frequency modulation to ensure an efficient transfer from the first-stage MOT. Subsequently, at lower intensity and without frequency modulation, the atoms are cooled further and captured by a red-detuned optical lattice operating at $\lambda \approx 813$~nm, the magic wavelength~\cite{kat03} of the clock transition. The depth of the optical lattice is typically $k_{\rm B} \times 14~\mu$K or 80~$E_{\rm rec}$, where $E_{\rm rec}=h^2/(2m\lambda^2)$ is the recoil energy from a lattice photon, $k_{\rm B}$ is the Boltzmann constant, $m$ is the mass of the atom and $h$ is the Planck constant. The horizontally-oriented lattice is formed by retro-reflection of the $\sim$300~mW output of a Ti:sapphire ring laser that is delivered to the position of the atoms via a large mode area (LMA) fibre to avoid Brillouin scattering at high power and focused to a waist radius ($1/\rm{e}^2$ intensity) of 45~$\mu$m. A portion of the laser output is sent to a frequency comb in order to monitor and stabilise its frequency. 

\begin{figure}
\centering
\includegraphics[width=0.6\columnwidth]{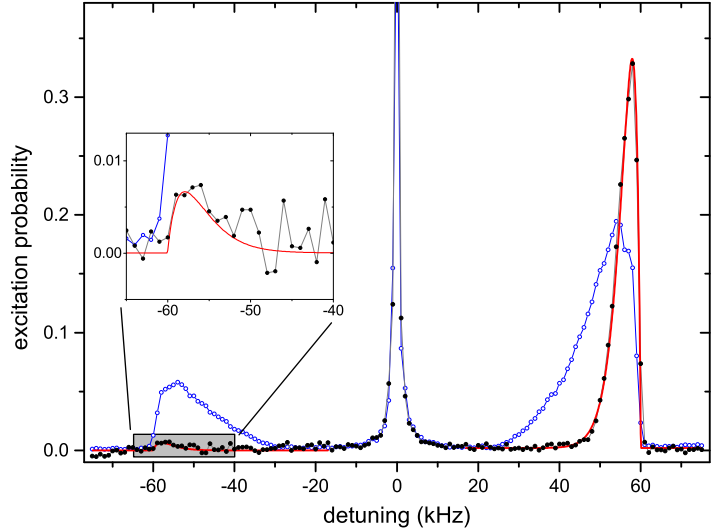}
\caption{Sideband spectra before (blue open circles) and after (black points) the energy filtering procedure as described in the text. Along with the filtered experimental data a simulated line shape (red line) adjusted to the experimental data is shown. The inset shows an expanded view of the expected position of the red axial sideband. Here, the amplitude of the simulated spectrum was chosen to be 0.02 times the amplitude of the blue sideband. The filtering procedure reduces the average excitation in the axial and radial directions, as seen in the increased ratio of blue to red sideband intensity and reduced width of the sidebands, respectively.}
\label{fig:sideband}
\end{figure}

Once the second stage MOT is switched off, the lattice-trapped atoms are spin-polarised by optical pumping into one of the stretched levels $m_F-9/2$ or $m_F=+9/2$. Then, the coldest atoms are selected by an energy filtering procedure wherein the optical lattice depth is ramped down within 5~ms to 28~$E_{\rm{rec}}$ for 20~ms and only the atoms with energies below this barrier are kept. After ramping the depth back to 80~$E_{\rm{rec}}$ in 5~ms, approximately 25~\% of the atoms remain in the trap. Here, we make use of an intensity control feedback loop acting on the diffraction efficiency of an acousto-optic modulator (AOM) between the laser and the LMA fibre. We servo the power of the retro-reflected light that travelled back through the LMA fibre as measured by a photodiode. The temperature of the ensemble is measured using sideband spectra (Fig.~\ref{fig:sideband}), which reveal an axial temperature $\lesssim 1~\mu$K and a radial temperature of 2~$\mu$K after the filtering procedure. This analysis is based on the line shape model developed in~\cite{bla09a}. The final temperature is insensitive to changes of the settings of several 10\%. The reduced atomic kinetic energies after the filtering procedure are beneficial for reducing systematic effects (site-to-site tunnelling and lattice light shifts) as discussed below. 

In order to combat impurities in the spin-polarisation (i.e. residual population in the other $m_F$ states), we apply a purification sequence as follows: First, a magnetic field of 625~$\mu$T is created utilizing the MOT coils in Helmholtz configuration. This field is sufficient to spectrally resolve the different $m_F$-components on the $\Delta m_F=0$ clock transition (about 675~Hz between neighbouring components). We apply a short clock laser pulse and observe a line width of 23~Hz of the individual components, which is the Fourier width of the 35~ms $\pi$-pulse. The clock laser beam is linearly polarised parallel to the applied magnetic field. Atoms remaining in the ground state are driven out of the lattice using $^1$S$_0$~--~$^1$P$_1$ resonant light. Depending on the initial spin-polarisation and the laser detuning the atoms are prepared in either the $m_F=+9/2$ or the $m_F=-9/2$ substate of $^3$P$_0$. 

The clock laser system at 698~nm is an extended cavity diode laser with an injection locked laser diode for power amplification. It is locked to a highly stable external resonator made from ULE (ultra low expansion glass). The frequency instability of the laser is further reduced by phase locking via a femtosecond comb to a 1.5~$\mu$m-fibre-laser stabilised to a single-crystal silicon cavity~\cite{hag13}.\footnote{This frequency comb is also part of the system counting the frequency of the clock laser.} The optical path lengths to frequency comb and optical lattice are actively stabilised utilizing AOMs~\cite{fal12}. With this system, the short-term laser stability allows for interrogation times of 320~ms without loss of spectral contrast with a Fourier limited transition line width of 2.5~Hz (full width at half maximum). For this measurement, the magnetic field is reduced to about 25~$\mu$T. The AOM used for pulsing sets the required intensities, both for preparation and interrogation pulses. This interrogation de-excites atoms from the $^3$P$_0$ state to the $^1$S$_0$ ground state on a $\pi$-transition. Even narrower lines are observed as shown in Fig.~\ref{fig:scheme} but for experimental robustness we operated the stabilisation with 2.5~Hz linewidth.

The average transition probability of the ensemble of about 1500 atoms is then detected destructively by collecting laser induced fluorescence from the 461~nm transition on a photomultiplier tube. A first pulse measures the population in the $^1$S$_0$ state. Next, the atoms remaining in the $^3$P$_0$ state are optically pumped to the $^1$S$_0$ state using two repump lasers, after which a second detection of the $^1$S$_0$ atoms reveals the number of atoms that did not make the clock transition. Finally, the background signal is measured with a third pulse, completing the experimental cycle.\footnote{In the fluorescence detection, we have replaced a photodiode used previously by a photomultiplier (Hamamatsu H11526-110-NF) for noise reduction.}

To realize a frequency standard, the clock laser is locked to the average frequency of the two $|m_F|=9/2$ components removing the linear Zeeman effect. To derive an error signal, four interrogations at the transition half-width points are performed: two for $m_F=-9/2$ and two for $m_F=+9/2$. The addressing of the four points is done by frequency offsets on the AOM for pulsing the clock laser light. Hence the laser frequency as seen by the frequency comb is unaffected by this addressing.

\section{Systematic frequency shifts}
\label{sec:srEffects}
There are a number of possible frequency shifts of the clock transition that must be considered when characterizing a lattice clock's accuracy. The most significant are the blackbody radiation shift and lattice light shifts. We will discuss these in detail while keeping the discussion of other effects brief. A summary of the uncertainty contributions is given in Tab.~\ref{tab:budget}.

\subsection{Blackbody radiation shift}
The largest frequency correction and residual uncertainty is due to the differential Stark shift of the clock states from blackbody radiation (BBR) emitted by the apparatus surrounding the atoms. In an earlier work we have reported the BBR shift as a function of the environmental temperature~\cite{mid12a} $T$ in first-order approximation scaling with $T^4$. The determination of this temperature is discussed here. Assuming that the BBR seen by the atoms is that of an opaque system, this task is reduced to finding the representative temperature of the system. The assumption is justified as the windows of the vacuum system are not transparent for room temperature BBR. The oven is an external heat source and will be treated separately below. Even without any knowledge of the emissivity of the materials or detailed knowledge of the temperature distribution, it is clear that the representative temperature will be between that of the hottest and the coldest point of the closed system. We therefore identified the points of extreme temperatures on our vacuum chamber and monitored these temperatures with Pt-100 sensors, which have an uncertainty of 0.04~K near room temperature. We have verified the specified uncertainty by calibration of individual sensors and inter-comparison of all sensors before and after thermal and mechanical shock testing. To minimize the interval of the temperature, heat sources inside or close to the vacuum system must be avoided. The Zeeman slower is cooled with water whose temperature is controlled to within 0.1~K. A second closed cooling system controls the temperature of the MOT magnetic field coils made from copper tubing, which are located inside the vacuum chamber and are also used to create the bias magnetic field used during state preparation and clock interrogation. The thermal capacity of these coils averages the time-dependent heat load throughout the experimental cycle. We monitor the temperature of the inlet and outlet water with additional Pt-100 sensors. We have verified with a thermal camera viewing through a (usually blocked) zinc selenide viewport that the temperature gradient measured by these sensors corresponds to the width of the oberserved temperature range on the surface of the coils. 

Once the minimal and maximal temperature $T_{\rm min}$ and $T_{\rm max}$ of the system are determined, a representative temperature and an associated uncertainty need to be derived. Without further knowledge or modelling of the thermal system, we assume a rectangular probability distribution for the representative temperature between $T_{\rm min}$ and $T_{\rm max}$. Hence, according to BIPM's \lq GUM: Guide to the Expression of Uncertainty in Measurement\rq~\cite{gum08}, $T=(T_{\rm min}+T_{\rm max})/2$ is the representative temperature with an uncertainty of $(T_{\rm max}-T_{\rm min})/ \sqrt{12}$, as this is the square root of the variance of the assumed probability distribution.

Compared to our previous frequency measurement~\cite{fal11}, we improved the thermal homogeneity across the vacuum chamber by reducing the thermal loads. Specifically, magnetic field compensation coils are now externally mounted and water cooled. Moreover, the current in the Zeeman slower magnet is switched off when the slower is not in use. Typical temperatures during operation have been $T_{\rm min}=293.71$~K and $T_{\rm max}=295.49$~K, and with the correction coefficients from \cite{mid12a} the shift correction is therefore $492.4(36) \times 10^{-17}$.

Despite these efforts, the uncertainty in the temperature is still a significant contribution to the uncertainty. One simple way to reduce this uncertainty is the inclusion of an additional time interval into the clock cycle, in which all magnetic coils are switched off. With an additional dead time of 1~s we have reached temperature intervals of $T_{\rm max}-T_{\rm min}$ of 1.2~K, corresponding to an uncertainty of $2.6\times 10^{-17}$ in studies following the absolute frequency measurement. This reduction of the heat load comes at the cost of the stability of the frequency standard, e.g. to $1.5\times 10^{-15} \sqrt{{\rm s}/\tau}$ for increasing the experimental cycle from 0.8 to 1.8 seconds.

The oven has no direct line of sight with the optical lattice but BBR photons emitted by the oven may be scattered towards the optical lattice. To consider the effect of these photons on the clock frequency, we apply a model that takes into account the temperature of the oven, a simplified geometry of the vacuum chamber, and the emissivity of the surface of the main vacuum chamber. For the uncertainty estimation of the oven induced BBR shift, uncertainties are applied to these parameters: The oven temperature is determined to $T_{\rm oven}= 770(50)$~K. Three assumptions are made: First, we model the chamber geometry as a sphere with radius of 17~cm and the representative temperature~$T$ determined as described above. A small fraction of the sphere's surface, which corresponds to the size of the aperture between oven and Zeeman slower tube (circle of 2~mm diameter), is replaced by a surface that emits a different radiation spectrum. Second, we assume that this spectrum is a superposition of a BBR spectrum of the oven at $T_{\rm oven}$ transmitted through the Zeeman slower and the BBR spectrum of the slower itself at~$T$.  Without detailed modelling, a rectangular distribution between 0 and 1 is assumed for the transmission probability of oven BBR photons, leading to a scaling of the amplitudes of both spectra by 0.50(29). As a third assumption we use the model of~\cite{mid11} to describe the effect of multiply scattered oven-BBR photons using inside the spherical chamber an emissivity of polished stainless steel of~0.1 (with assumed minimal and maximal values of 0.07 and 0.2)~\cite{wie79}. The total effective shift due to scattered BBR photons emitted by the oven is $-1.6\times 10^{-17}$. Considering the uncertainties of the three discussed parameters, we obtain a combined uncertainty of $1.2\times 10^{-17}$.

\subsection{Optical lattice light shifts}
A key concept of optical lattice clocks is to trap the atoms in a light field that, to first order in intensity, perturbs the two clock states equally. This is achieved by tuning the frequency of the laser light that creates the optical lattice to a magic wavelength; for Sr, the red-detuned magic wavelength is near 813~nm. Even at shallow trap depths, small effects due to higher order light shifts (both multipolar and two-photon) must be considered, as pointed out in \cite{wes11}. We will account for these as discussed below, but first we describe the measurement of the magic wavelength.

\begin{figure}
\centering
\includegraphics[width=0.6\columnwidth]{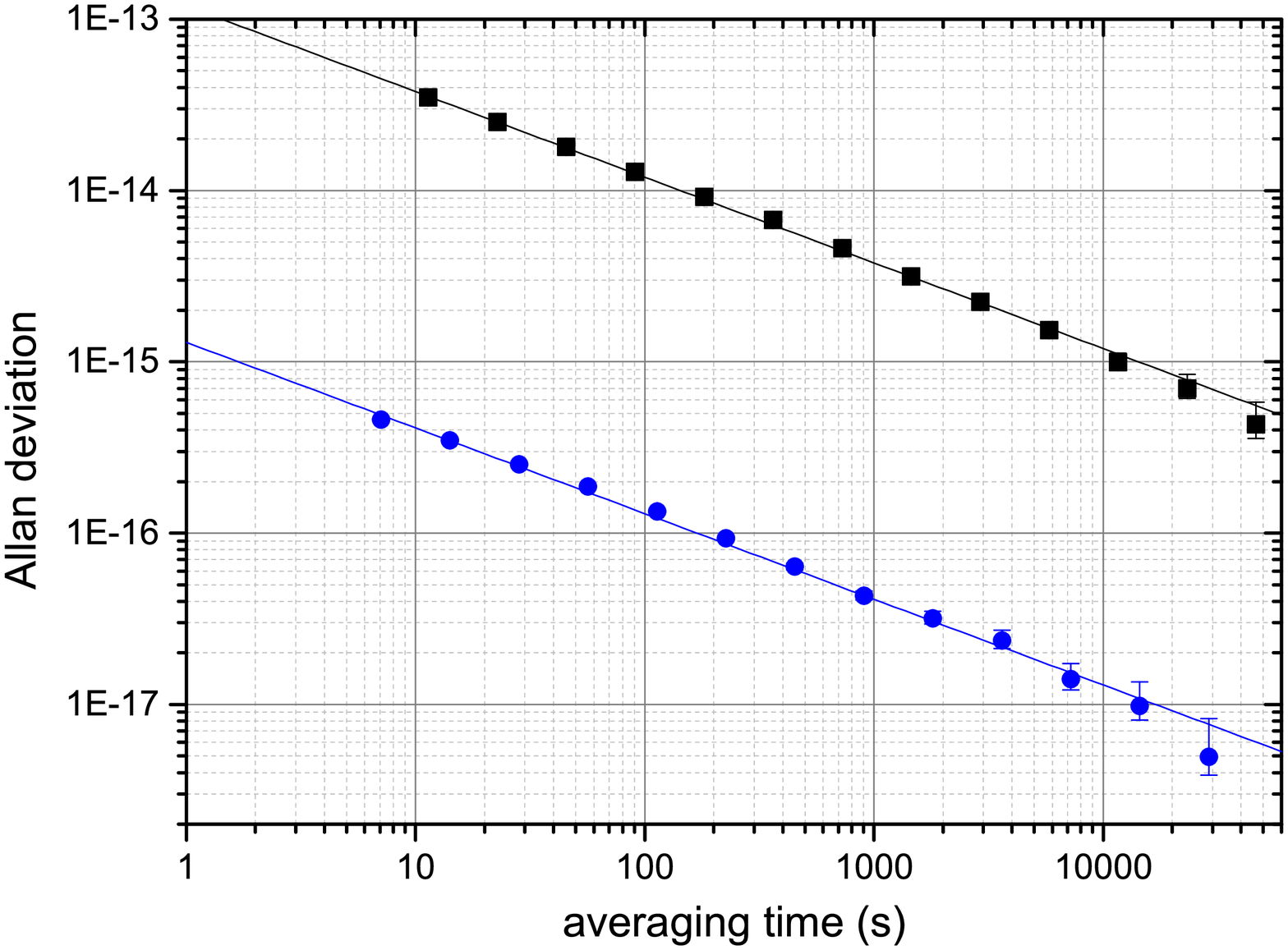}
\caption{Instabilities of the frequency measurement. Shown are the Allan deviations of the comparison of the Yb$^+$ standard (black squares) and CSF2 and the total Allan deviation of an interleaved stabilisation of the Sr frequency standard (blue circles). The estimated instability of the Sr frequency standard is a factor of two smaller than this observation. The lines indicate instabilities of 1.2$\times 10^{-13} \; ({\rm s}/\tau)^{1/2}$ and $1.3\times 10^{-15} \; ({\rm s}/\tau)^{1/2}$.}
\label{fig:stability}
\end{figure}

For a particular lattice frequency, we compare the frequency of the clock transition with two different lattice intensities (80~$E_{\rm{rec}}$ and 160~$E_{\rm{rec}}$) using an alternating stabilisation method~\cite{lis09,deg05a}. The difference of the two offset frequencies between the atomic transition and the clock laser cavity mode is measured with independently working digital servo loops and gives the clock frequency dependence on the lattice intensity. The total Allan deviation~\cite{how99} shown in Fig.~\ref{fig:stability} (blue markers) indicates the stability of this interleaved measurement as $\sigma_y \left( \tau \right) \approx 1.3\times 10^{-15} \left (\rm{s}/\tau\right)^{1/2}$. We expect that the instability of the clock operated with a single stabilisation instead of an interleaved one is at least a factor of two smaller. Although the atoms are very cold after the filtering procedure, their population distribution over the trap levels as observed in sideband spectroscopy needs to be considered when deriving the average lattice light intensity seen by the atoms~\cite{bla09a}. We then correct the observed frequency difference for a difference in the quadratic light shift (hyperpolarisability) according to \cite{wes11} using the coefficient given in~\cite{let13} and the averaged intensity. This measurement and analysis is repeated at nine lattice frequencies in order to identify where the shift crosses zero. We derive the first order Stark-shift cancelling wavelength for $m_F=\pm 9/2$ with a linear regression of the corrected frequency differences normalized to the average light intensity experienced by the atom due to thermal motion and find $368\,554\,465.0$~MHz. 

In the measurement of the first order Stark-shift cancelling wavelength, uncertainties due to the hyperpolarisability, residual tunnelling, and collisions (discussed below) are added in quadrature along with the statistical uncertainty to derive the uncertainty contribution due to the scalar and tensor shifts as $9\times 10^{-18}$. The associated uncertainty of the trapping laser frequency is 3~MHz. Note that the first order Stark-shift cancelling frequency depends not only on the chosen $m_F$ state but also on the angle between the polarisation of the lattice light and the magnetic field; in our case these are collinear. When comparing to the values for the magic wavelength published in~\cite{let13, wes11} our value needs to be shifted by +286 MHz because of the tensor shift, and good agreement is found.

We now consider corrections and uncertainty contributions due to effects that are non-linear in the lattice intensity. First, the hyperpolarisability correction at the operational lattice depth of 68~$E_{\rm rec}$ (thermally averaged) during the frequency measurement is $-5\times 10^{-18}$ with an uncertainty of $2\times 10^{-18}$, where we have allowed for 20 percent uncertainty in the intensity seen by the atoms and include the uncertainty of the published correction coefficient of $0.45(10)\ \mu{\rm Hz}/E_{\rm rec}^2$ \cite{let13}. Second, higher order interactions (magnetic dipole and electric quadrupole \cite{tai08}) have not been observed, but an upper limit was derived in~\cite{wes11}. The shift is smaller than $\pm 0.31\;{\rm mHz}/\sqrt{E_{\rm rec}}$ leading to an uncertainty contribution of $6\times 10^{-18}$.

Another uncertainty contribution related to the optical lattice arises from atoms tunnelling from site to site~\cite{lem05}, thus introducing a first-order Doppler shift. In another picture, the periodicity of the trapping potential leads to energy bands that are not necessarily equally populated, which could cause line shifts on the order of the bandwidth. For a perfectly horizontal lattice with 80~$E_{\rm{rec}}$ depth, the width of the two lowest energy bands are 13~mHz and 826~mHz. According to~\cite{sia08}, energy differences between sites lead to a suppression of the tunnelling, i.e., the bandwidths are reduced. The lattice is nearly horizontal with a small tilt of 2.7~mrad, leading to an energy difference of 2.4~Hz between two neighbouring sites due to gravity. The experimental observations of~\cite{sia08} support the reduction of the bandwidth by at least a factor of fifty for our parameters. We apply this and consider the population of the vibrational bands as derived from sideband spectra (see Fig.~\ref{fig:sideband}), i.e., fewer than two percent of the atoms are in the first axially excited band while no population remains in the second band due to the filtering, and we estimate a maximal fractional frequency shift due to tunnelling to be $1\times 10^{-18}$.

\subsection{Clock laser}
Although low in intensity, the clock laser light may cause a frequency shift of the clock transition due to coupling to other states. With a measurement of the intensity and the sensitivity of $-13(2){\rm Hz}/({\rm W}/{\rm cm}^2)$ from \cite{bai07} we determine that a light shift of less than $5\times 10^{-19}$ is caused by the clock laser, to which we assign an uncertainty of $1\times 10^{-18}$.

A second uncertainty related to the clock laser needs to be considered: a drift of the clock laser reference cavity may lead to a systematic shift as the sides of the line are probed in the same order for each frequency correction. This effect is largely removed by drift-compensation with a linearly ramped AOM frequency. The residual systematic error is calculated to be $3\times 10^{-19}$ based on the update rate of the drift estimate (gain of $\beta=0.02$ per correction) and the maximal observed change of the linear drift ($1.1\times 10^{-7}$~Hz/s$^2$) according to equation~(5) in \cite{fal11}.

\subsection{Transfer of clock laser light and Doppler effect}
The clock laser light is transferred via optical fibres both to a frequency comb and to the Sr atoms. Both transfers use single mode, polarisation-maintaining fibres and optical path length stabilisations. The uncompensated optical path is kept below 0.5~m. While the path length stabilisation for the link to the comb is active at all times, the stabilisation of the path to the atoms is operated only during the interrogation pulses. The transient behaviour during switching was identified as possible error-source. We investigated this as described in \cite{fal12} and obtain an uncertainty contribution of the optical path length stabilisation of $8\times 10^{-19}$ under present experimental conditions.\footnote{We benefit from longer clock pulses compared to previous work due to a smaller clock laser instability.} Since the mirror producing the standing wave of the optical lattice is used as reference for the remote fibre end, first order Doppler shifts due to lattice vibrations or AOM chirps due to heating are removed. Second order Doppler shifts are not significant here.

The same AOM used for pulsing the clock laser and optical path length stabilisation also provides the shift from the line centre to the desired $m_F$ component and the chosen side of the probed line. The power of the diffracted light depends on the frequency of the AOM, which in principle leads to asymmetric excitation probabilities at the low and high frequency halfwidth points that would be incorrectly interpreted as line shifts. However, this effect is negligibly small.

\subsection{Cold collisional shifts}
Several atoms occupy each lattice site and collisions may shift the resonance frequency~\cite{nic12}. We have measured the density-dependent frequency shift by interleaving interrogations with few and many Sr atoms, which is realized by different loading times of the first stage MOT. The observed frequency shift is consistent with zero. A linear scaling (see e.g.~\cite{sor00}) to the atom number used in our frequency measurement results in an uncertainty of $5\times 10^{-18}$. Typically the clock is operated with about 1500 atoms.

\subsection{Collisions with background gas}
We follow \cite{gib13} to relate expected frequency shifts due to collisions with background gas atoms and molecules to observed trap lifetimes. We estimate the particle density for a number of gases (H, H$_2$, He and Xe) to reproduce our observed trap lifetime of 7~s (last term in Eq.~(3) in~\cite{gib13}). We then determine the expected frequency shift using the long-range coefficients $C_6$ for both clock states.\footnote{The approximation using $\Delta C$ made in \cite{gib13} was undone to account for differences of the long-range coefficients of typically 30 percent.} The ratio of $C_6$ coefficients of the two clock states varies weakly between typical background gas species~\cite{mit10a}, and we finally assign a correction of $1.5(15)~\times 10^{-18}$.

\subsection{Line pulling}
After the spin-polarisation and purification steps, the population is almost entirely in the chosen $m_F$ state, but residual populations in other $m_F$ states may persist. Transitions driven from neighbouring $m_F$ components will lead to a line pulling by creating an asymmetric background on the observed line. We derive experimentally an upper limit of five percent for atoms in other states, and with the known detuning of the $\pi$-transition of $m_F=\pm 7/2$, the line pulling effect is found to be less than $2\times 10^{-19}$.

If the polarisation of the laser is not parallel to the quantization axis set by the magnetic field, i.e. due to birefringence of the viewports or a tilt of the polarisation axis, another line pulling effect may occur: instead of a $\pi$-polarisation coupling scheme with just the two $m_F=\pm 9/2$ levels, the admixing of sigma polarisation couples to the $m_F=\pm 7/2$ state. In this $\Lambda$-scheme the $\pi$ and $\sigma$-transitions are driven coherently, in which case an incoherent superposition of line profiles is not an adequate description. Instead, we numerically integrated the time-dependent Schr\"odinger equation of the three-level system and obtained an upper limit for the frequency shift of $2\times 10^{-19}$, where we assumed a fractional power in the $\sigma$-polarisation of less than eleven percent. This upper limit is derived from the minimal signal of a $\sigma$-transition that would have been visible on scans of the expected position of that transition. These two line pulling effects combined contribute an uncertainty of $3\times 10^{-19}$ fractional frequency shift.

\subsection{Zeeman shifts}
Stabilising the clock laser to the average frequency of the two Zeeman components not only removes the linear Zeeman shift and the vector light shift (the magnetic field stability over subsequent interrogation cycles is high enough) but also provides an estimate of the total magnetic field experienced by the atoms, including any stray fields provided that the lattice vector shift is small~\cite{wes11}. This enables a correction for the quadratic Zeeman shift, which is $-3.2\times 10^{-17}$. The associated uncertainty of $1.3\times 10^{-18}$ is mostly due to the uncertainty of the magnetic field as determined from the splitting of the clock transitions $m_F=-9/2$ and $+9/2$ of 240~Hz, while the correction coefficient contributes with negligible uncertainty ($2\times 10^{-19}$). During measurement of the magic wavelength, we measured the vector light shift to be smaller than 0.2~Hz, which can be neglected in the quadratic Zeeman shift correction.

\subsection{DC electric fields}
Residual static electric fields, e.g. from contact potentials, may be present that shift the frequency of the clock transition as observed in~\cite{lod12}. We have measured stray electric fields by applying additional electric fields and detecting the change of the transition frequency as a function of the applied field. The Stark shift is proportional to the square of the total electric field and hence a parabolic behaviour is observed. The projection of a residual electric field on the applied field is determined by the offset of the parabola from zero applied field strength. We have measured three projections of the stray field by applying moderate voltages to electrodes near the optical lattice. The known DC polarisability allows us to infer the applied field strength from the observed shift and applied voltage. 

In order to derive the absolute value of the residual electric field from these three projections, the three angles between the applied fields must be determined. This is done by comparing the applied field strength with the observed projection of this field, which is seen in the parabola offset while generating a known electric field with one of the other two electrodes (much stronger than the residual field). With these measurements the residual electric field was found to be 15.8(22)~V/m. The observed field is likely due to a voltage difference of about 0.7~V between the two coils providing the bias magnetic field, which is mostly caused by a protection diode between the coils.

During the frequency measurement, however, the electric field was larger because the magnetic field coils had not been operated near ground potential but rather at a potential of about 21~V, while the nearby capacitor used in \cite{mid12a} was grounded. With hindsight, we measured the frequency shift with an interleaved stabilisation, alternating the voltage on the coils between 21~V and ground. The observed shift of $-1.52\times 10^{-16}$ is consistent with the expected electric field strength and is included as a correction in the uncertainty budget. The associated uncertainty ($3.3\times 10^{-17}$) is mostly due to instabilities of the potential of the coils of $\pm 2$~V. We now operate the experiment with the coils near ground potential to remove this considerable uncertainty.

\begin{table}
\centering
\begin{tabular}{lr@{}lr@{}l}
\hline
{\bf effect} &
\multicolumn{2}{c}{\bf correction} & 
\multicolumn{2}{c}{\bf uncertainty} \\ 
 &
\multicolumn{2}{c}{($10^{-17}$)} & 
\multicolumn{2}{c}{($10^{-17}$)} \\ 
\hline
\hline
BBR room &
492&.4 &
3&.6\\
BBR oven &
1&.6 &
1&.2\\
second-order Zeeman &
3&.2 &
0&.13\\
cold collisions &
0 & &
0&.5\\
background gas collisions&
0&.15 &
0&.15\\
line pulling &
0 & &
0&.03\\
lattice scalar/tensor &
\hspace{5ex}0 & &
\hspace{5ex}0&.9\\
hyperpolarisability &
$-0$&.5 &
0&.2\\
lattice E2/M1 &
0 & &
0&.6\\
tunnelling &
0 & &
0&.1\\
probe light &
0 & &
0&.1\\
optical path length error &
0 & &
0&.08\\
servo error &
0 & &
0&.03\\
DC Stark shift &
15&.2 &
3&.3\\
\hline
{\bf total} &
512&.0 &
5&.2\\
\hline
\end{tabular}
\caption{Corrections and uncertainties of the strontium frequency standard during the frequency measurement given in fractional frequencies.}
\label{tab:budget}
\end{table}

\section{Absolute frequency measurement}
\label{sec:absfreq}
The frequency of the stronium lattice clock was measured against the two caesium fountain clocks CSF1 and CSF2 operated at PTB \cite{wey11,wey01,wey01a,ger10}, which are among the best primary frequency standards. During the measurement, an ytterbium ion clock based on the octupole transition $^2$S$_{1/2}$($F$=0)--$^2$F$_{7/2}$($F$=3) of $^{171}$Yb$^+$ \cite{hun12,hun12a} was used as a flywheel oscillator, which allows us to bridge gaps in the operation of the Sr clock and extend the usable averaging time versus the Cs fountains. To improve the stability of CSF2 we utilized the Yb$^+$ clock laser to transfer its stability to the microwave regime with the help of a fs-laser frequency comb~\cite{lip09,wey09,tam14}.

\begin{figure}
\centering
\includegraphics[width=0.8\columnwidth]{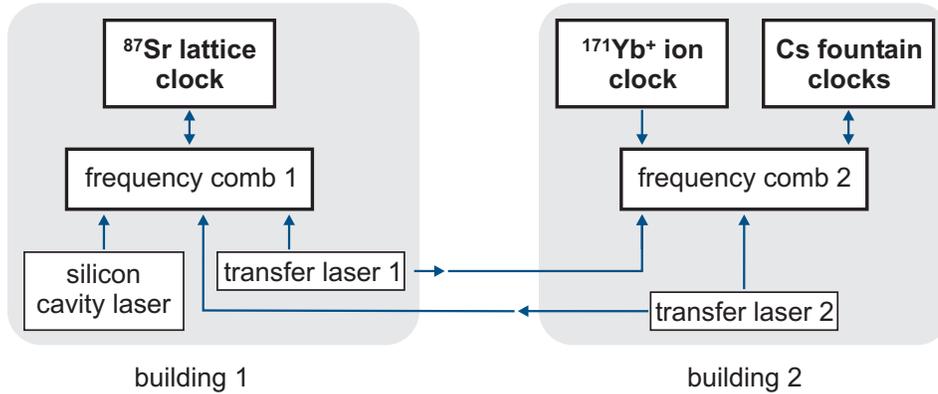}
\caption{Overall experimental setup of the frequency measurement. Two frequency combs are employed to connect two optical clocks and the Cs fountain clocks as well as for the transfer of frequency stability. The connection between these combs, i.e., between two buildings separated by 300~m, is established by stabilised optical fibre links.}
\label{fig:network}
\end{figure}

The overall setup of the frequency measurements is shown in Fig.~\ref{fig:network}. We employ two frequency combs, one in each building. Two ultrastable cw-laser at 871~nm and 1543~nm link these two combs. This twofold connection establishes a loop geometry that enables the characterization of this transfer scheme with respect to accuracy, stability, and reliability. Moreover, stabilised optical links between the buildings are preferable to traditional microwave links because they are less susceptible to thermally-driven phase shifts that caused a significant uncertainty contribution in our previous measurement~\cite{fal11}.

Most of the optical paths of the measurement set-up are actively stabilised. The residual few meters of yet unstabilised optical path are passively shielded against environmental perturbations and their length variations influence the measurement. Based on a loop measurement with the two transfer lasers and the two combs, we derive an uncertainty contribution due to the frequency comparison in the optical regime of less than $1\times 10^{-18}$. To ensure this accuracy, the synchronization of the dead-time free $\Pi$-counters in the two buildings is established via a pulse-per-second signal, and the propagation time ($< 1~\mu$s) is the only known source of a temporal offset between the gates of the counters at both frequency combs. Gate synchronization is important to ensure proper noise correlation and to avoid frequency offsets due to frequency drifts of the transfer lasers. To detect possible cycle slips in the phase tracking of the signals we split each beat signal and record it twice after independent filtering and phase tracking. A differential phase offset of 2$\pi$ is taken as indication for a cycle slip and led to an exclusion of the data point.

The data analysis to derive the frequency of the strontium clock is performed in two steps:
\begin{itemize}
\item The frequency of the Yb$^+$ clock was measured relative to the Cs fountains for a total time of approximately $350\,000$ seconds between the 10$^{\rm th}$ and 18$^{\rm th}$ of December 2012. This leads to statistical uncertainties of $2.5\times 10^{-16}$ and $2.0\times 10^{-16}$ due to the instabilities of the Cs fountains and the observed white-noise $1/\sqrt{\tau}$~dependence of the frequency measurement (see Fig.~\ref{fig:stability}). The systematic uncertainties of the Cs fountains are $7.3\times 10^{-16}$ for CSF1 and $4.0\times 10^{-16}$ for CSF2. Averaging the two obtained frequency values leads to an uncertainty of $3.88\times 10^{-16}$.
\item In the same period, the frequency ratio of the strontium standard and the Yb$^+$ standard was measured for a total of approximately $165\,000$ seconds. We observe a clean $1/\sqrt{\tau}$~characteristic of the total Allan deviation in this comparison, leading to a statistical uncertainty of $1.3\times 10^{-17}$ for the comparison of the Yb$^+$ and the Sr clock.
\end{itemize}
Because both measurements show white frequency noise dependence, we can use the longer measurement time of the Yb$^+$ clock against the Cs fountain clocks to improve the statistical uncertainty of the Sr--Cs measurement, while not requiring a systematic frequency shift evaluation of the Yb$^+$ clock. This procedure leads to an absolute frequency measurement of the Sr clock transition with a relative frequency uncertainty of $3.9\times 10^{-16}$ (see Tab.~\ref{tab:absolute}). We included the correction of the gravitational red shift due to the height difference between the Cs fountains and the Sr clock, which was determined in a levelling measurement with 6~mm uncertainty.

\begin{table}
\centering
\begin{tabular}{lr@{}l}
\hline
{\bf effect} &
\multicolumn{2}{c}{\bf uncertainty} \\ 
 &
\multicolumn{2}{c}{($10^{-17}$)} \\ 
\hline
\hline
comparison Cs and Yb$^+$ including realization of the SI second&
38&.8\\
statistics of Yb$^+$ and Sr comparison&
1&.3\\
Sr systematics &
\hspace{5ex}5&.2\\
transfer scheme&
0&.1\\
gravitational red shift&
0&.07\\
\hline
{\bf total} &
39&\\
\hline
\end{tabular}
\caption{Uncertainties of the absolute frequency measurement given in fractional frequencies.}
\label{tab:absolute}
\end{table}

\section{Results and discussion}
\label{sec:results}
\begin{figure}
\centering
\includegraphics[width=0.6\columnwidth]{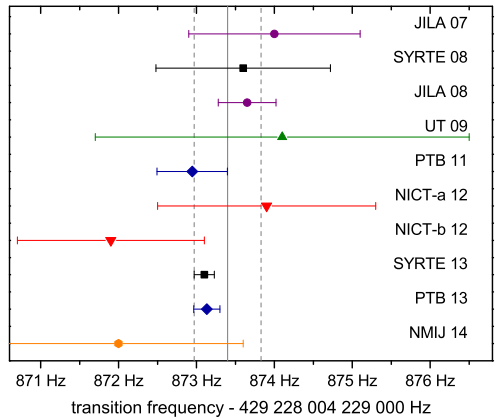}
\caption{Comparison of measured absolute frequencies of the 5s$^2$~$^1$S$_0$~--~5s5p~$^3$P$_0$ transition in $^{87}$Sr. The frequencies are taken from these publications: JILA~07~\cite{boy07}, SYRTE~08~\cite{bai08}, JILA~08~\cite{cam08b}, UT~09~\cite{hon09}, PTB~11~\cite{fal11}, NICT-a~12~\cite{yam12}, NICT-b~12~\cite{mat12}, SYRTE~13~\cite{let13} and NMIJ~14~\cite{aka14} while PTB~13 is obtained in this work. The vertical line indicates the recent recommendation by the BIPM in 2013~\cite{cip13} while the dashed lines show the assigned frequency uncertainty of $1\times 10^{-15}$. The frequencies and uncertainties from data published before 2013 are not updated with improved knowledge on the blackbody correction coefficients~\cite{mid12a,saf13}.}
\label{fig:frequencies}
\end{figure}

We have measured the absolute frequency of the $^{87}$Sr clock transition to be $429\,228\,004\,229\,873$.13(17)~Hz. This result agrees well with previous measurements and the recommended value by BIPM as shown in Fig.~\ref{fig:frequencies}. In particular, we find excellent agreement between our result and the recent high-accuracy measurement from LNE-SYRTE~\cite{let13}, demonstrating one of the best levels of agreement between two independent measurements of the same physical quantity.

With proper grounding of the magnetic field coils, the DC Stark shift and its uncertainty is largely removed. In combination with the demonstrated uncertainty in the BBR shift correction due to reduced thermal load, the Sr frequency standard is now operated with a systematic uncertainty of $3\times 10^{-17}$. Progress in reducing the BBR uncertainty further will likely require improved control over the thermal environment \cite{mid11} or direct measurement of the radiation field~\cite{blo13}. If the lattice clock is operated at room temperature, the accuracy is currently limited by the knowledge of the atomic reaction to the BBR field, in particular the so-called dynamic contribution \cite{mid12a, saf13}, to an uncertainty of about $4 \times 10^{-18}$. Though this correction can in principle be quantified better by a measurement of dipole matrix elements as demonstrated for Yb \cite{bel12}, we believe that a reduction of the shift itself by lowering the environmental temperature is the more promising approach to reach uncertainties of $10^{-18}$ and below.

Any future design of our Sr standard aiming at a further reduction of the uncertainty should not only address the BBR shift but also orient the lattice along gravity~\cite{lem05}: Given the current knowledge of the shift coefficients \cite{wes11}, higher order effects in the lattice light intensity will limit the total uncertainty at the $10^{-17}$ level in our apparatus because of the relatively deep trap needed to levitate the atoms against gravity in a horizontal lattice. 

Direct comparisons of optical clocks already provide smaller relative uncertainties~\cite{ros08, cho10, hin13, blo13} in the frequency ratio than can be achieved with absolute frequency measurements. Nonetheless, additional absolute frequency measurements such as the one described here are needed to provide a solid basis for a discussion of a re-definition of the SI second.

\section*{Acknowledgements}
The authors would like to thank Heiner Denker, Nico Lindenthal, and Ludger Timmen (Institut f\"ur Erdmessung, Leibniz Universit\"at Hannover) for the levelling of the clocks, Thomas Legero (PTB) for providing a transfer laser, and Kurt Gibble (Department of Physics, The Pennsylvania State University) for helpful discussions.

This work was supported by the Centre of Quantum Engineering and Space-Time Reseach (QUEST), the German Research Foundation (DFG) through RTG 1729 \lq Fundamentals and Applications of ultra-cold Matter\rq, and the projects \lq International Timescales with Optical Clocks\rq\ (ITOC), \lq New Generation Frequency Standards for Industry\rq\ (IND14), and SOC2. The ITOC and IND14 projects are part of the European Metrology Research Programme (EMRP). The EMRP is jointly funded by the EMRP participating countries within EURAMET and the European Union. SOC2 is a EU-FP7 project: \lq Towards Neutral-atom Space Optical Clocks: Development of high-performance transportable and breadboard optical clocks and advanced subsystems\rq\ (Project No. 263500).

\section*{References}
\bibliographystyle{unsrt}

\end{document}